\documentclass[runningheads]{llncs}
\usepackage[T1]{fontenc}
\usepackage{graphicx}
\usepackage{amsmath,amssymb,amsfonts}
\begin{document}
\title{Application of genetic algorithm to load balancing in networks with a homogeneous traffic flow}
\titlerunning{Application of genetic algorithm to load balancing...}
\author{Marek Bolanowski\inst{1}\orcidID{0000-0003-4645-967X} \and
Alicja Gerka \inst{2}\orcidID{0000-0001-7406-8495} \and
Andrzej Paszkiewicz\inst{1}\orcidID{0000-0001-7573-3856} \and
Maria Ganzha\inst{3}\orcidID{0000-0001-7714-4844} \and
Marcin Paprzycki\inst{3}\orcidID{0000-0002-8069-2152}}
\authorrunning{M. Bolanowski et al.}
\institute{Rzeszów University of Technology, Rzeszów, Poland \\
\email{\{marekb, andrzejp\}@prz.edu.pl}
\and
\email{alicja.gerka@op.pl}
\and
Systems Research Institute Polish Academy of Sciences, Warsaw, Poland
\email{\{maria.ganzha, paprzyck\}@ibspan.waw.pl}\\
}
\maketitle              
\begin{abstract}
The concept of extended cloud requires efficient network infrastructure to support ecosystems reaching form the edge to the cloud(s). Standard approaches to network load balancing deliver static solutions that are insufficient for the extended clouds, where network loads change often. To address this issue, a genetic algorithm based load optimizer is proposed and implemented. Next, its performance is experimentally evaluated and it is shown that it outperforms other existing solutions.

\keywords{extended cloud \and computer network \and load balancing \and SDN \and routing \and IoT \and self-adapting networks.}
\end{abstract}
\section{Introduction}
Today, typical ``data processing systems'' consist of ``distributed data sources'' and (a) ``central cloud(s)''. However, the advent of Internet of Things forces changes, since  the ``decision loop'', from the ``sensor(s)'' to the cloud and back to the ``actuator(s)'', may take too long for (near-)real time applications. This problem is addressed by, so called, \textit{Extended Cloud} (EC). Here, the whole ecosystem becomes a logical/virtual cloud. Obviously, the EC encompasses highly heterogeneous hardware (including networking devices). The typical approach to functionalize this vision is by application of Software Defined Networking (SDN). However, the question of SDN load balancing remains open  as, for networks with varying communication loads, \textit{dynamic} load balancing is needed.

Here, note that, in the classical approach, in both the L2 and L3 layer protocols (of the ISO/OSI model; STP), in the case of multiple connection paths, as a rule, only one is used, while the remaining paths become ``backup connections''. Obviously, multiple mechanisms to balance the load have been proposed, e.g.: by modifying the STP root priority parameters for individual VLANs; or by routing protocols attempting at balancing the load (OSPF equal-cost multi-path, EIGRP unequal-cost multi-path)~\cite{bl1,bl2,bl3}. However, since they redistribute the load ``locally'', some links remain overloaded, while others are not used.

In standard routing protocols, after a ``warm-uptime'', the final routing table is formed. For instance in the RiverBed Modeler~\cite{bl4}, by default, the routing stabilization occurs after 300 seconds. Attempts to dynamically modify routing parameters are focused on modification of the routing algorithms, link weights, etc. (\cite{bl5,bl6}). However, their \textit{implementation potential} is low. For example, continuous modification of link weights, or direct interference with the routing protocol settings, during dynamic routing, may require reset of the Open Shortest Path First (OSFP) process. This, is very time consuming, and has negative impact on network reliability. Here, note that the keys to effective management of network load balancing are: (I)~global view of the network, and (II)~elimination of overloaded links through dynamic adaptation of the routing policy. This requires an ``arbitrator'', which will collect information about communication requests and dynamically manage routing (\cite{bl7,bl8,bl9}). At present, this vision is met by the SDN  paradigm, which also enables easy implementation of novel traffic control algorithms. This, in turn, allows balancing loads in the aggregation layer, connecting EC elements that, in addition to classic elements (servers, routers, etc.), may contain ``IoT elements'', e.g. sensors and actuators\cite{blm1}. Here, note that such elements are particularly sensitive to communication delays,  e.g., in remote control of industrial elements \cite{bl10,bl11,bl12} (ECs realizing Industry 4.0 scenarios). 

Separately, note that classic load balancing algorithms, when applied to the SDN-based networks, are NP-hard~\cite{wang_load-balancing_2018}. Therefore, a genetic algorithm-based approach (the SDNGALB algorithm), to SDN network load balancing, is being proposed. The proposed solution delivers self-adaptive network that can evolve over time, adapting to communication requests, e.g. within the EC running network. The proposed approach has been implemented and tested in lab-based network, showing its advantages. 

In this context, Section~\ref{section1} summarizes pertinent literature. Next, Section~\ref{section2} contains problem formulation, and the proposed algorithm. In Section~\ref{section3}, experimental results are presented and discussed. Section~\ref{section4} summarizes main contributions and suggests future work directions.

\section{Related work \label{section1}}

Effective optimization of use of network and system resources, in the Extended Cloud, requires effective load balancing. Since the nature of the problem is NP-hard, metaheuristic algorithms have been proposed. Therefore, in what follows, only results related to application of metaheuristics-based approaches to network load balancing will be discussed. Readers interested in a comprehensive overview of other load balancing approaches may consult~\cite{bl17,bl18,bl23,bl24,bl26}.

In~\cite{bl13}, the authors compared the effectiveness of metaheuristics-based approach to network resource and energy optimization. The performance of Particle Swarm Optimization, Ant Colony Optimization, Artificial Bee Colony, Genetic Algorithm (GA), and Bat scheduling algorithm, has been explored. These algorithms can be also used for load balancing in actual networks, but the cited work focuses only on virtualized distributed systems.

Work reported in~\cite{bl14} considers balancing the load of SDN controllers, using a genetic algorithm, to improve the operating parameters of the network. The authors have successfully used GA to pair elements of a cluster of controllers and controlled switches. Similarly, in~\cite{bl15}, research was focused on the impact of sudden changes in network flows on the load distribution, between SDN controllers. Such changes may impair processing of data flows by the most loaded controllers and, simultaneously, limit use of lightly loaded ones. Here, load balancing of other network elements, except for the SDN controllers, was not considered. 

In~\cite{bl19}, a genetic algorithm-based strategy for load balancing in SDN optical networks was proposed. This approach, however, focuses on optimal localization of new (added) links, rather than on balancing load in existing connections. 

A GA-based, strategy for balancing load of end devices, in wireless metropolitan area networks, to optimize the performance of mobile applications, was proposed in~\cite{bl20}. The solution deals with optimal distribution of computing tasks across the network infrastructure. It was found that while lowering the overall network response times, it does not directly affect the load on individual links.

Work described in~\cite{bl21}, is focused on SDN load balancing and uses genetic algorithm to search for the best path for a given traffic, taking into account the current load, bottlenecks and the shortest possible path. The proposed solution has been tested for Fat-Tree topology only. 

Optimization of weights, in MPLS and OSPF networks, is a separate research area. However, it is directly related to the topic discussed here. The search for a solution to the OSPF Weight Setting problem is very important in the context of the load balancing. It is also NP-hard, and the authors of~\cite{bl27} stated that even the search for an approximate solution should be treated as an NP-hard task. In~\cite{bl28}, the authors further confirm that solving the problem of load balancing is NP-hard, while being very important from the point of view of traffic engineering. Here, the authors take into account the matrix of communication requests between the nodes, in order to define the traffic, but the solution is dedicated to networks with a diameter of no more than 5. 

Authors of~\cite{bl29}, proposed a genetic algorithm to optimize the link weights used in the OSFP protocol, to minimize the maximum load on the links, perceived as the percentage saturation of the communication channel. This is a continuation of the work presented in~\cite{bl30}. Here, the authors propose to use GA to configure the OSPF weights for several popular network topologies, and the optimization goal of such a distribution of traffic is to minimize the total cost of the flow, which is defined as the quotient of link load and link capacity. Finally, in~\cite{bl31}, multi-objective particle swarm optimization algorithm is used to efficiently solve the OSPFWS problem. Here, scalar cost function, derived from optimization metrics such as maximum utilization, number of congested links, and number of unused links, is used in the proposed approach.

Overall, based on a comprehensive analysis of works related to use of standard, and metaheuristics-based, approaches to network load balancing, the following conclusions have been formulated:

\begin{itemize}

\item Use of metaheuristics-based approaches, to solve the load balancing problem in computer networks, remains an active (and promising) research area.

\item Solutions found in the literature are focused mainly on use of static values of weights for communication links. This leads to creation of a temporarily optimal, but static, connection structure, and leaves an open research gap.

\item In the conducted work, special attention should be paid to the possibility of applying the developed solutions in the environment of real network devices. Among other things, attention should be paid to reducing the time needed to find solutions, which has a direct impact on the time needed to reconfigure the network in response to changing operating conditions. 

\end{itemize}

To address limitations of found approaches, and to deliver solution applicable to ECs, a genetic algorithm, with high implementation potential, is proposed. To describe it, let us start from a formal problem description.

\section{Problem formulation and proposed solution \label{section2}}
In what follows, computer network will be represented by a directed graph $G(N,E)$, where $N$ is a set of nodes, representing network devices, and $E$ is a set of edges representing network links. Each edge ${{e}_{ij}}\in E$ is assigned a weight ${{w}_{ij}}$, the modification of which will affect the current shaping of the routing policy. 
Moreover, the following assumptions have been made:
\begin{itemize}
    \item Communication channels, represented as ${{e}_{ij}}\in E$, have the same bandwidth. Nevertheless, the application area of the algorithm includes: distribution layer of  networks, small and medium operator networks, MESH and IoT networks.
    
    \item $G\left( N,E \right)$ is a directed graph, which allows to take into account the asymmetry of flows, typical for routing protocols operating in real networks.  

    \item Network switches, routers and intermediary nodes are treated as ``identical network devices'' because, from the point of view of SDN network control, their distinction is irrelevant~\cite{bl32}.
    
\end{itemize}

The network topology is represented by a graph adjacency matrix $G(N,E)$, denoted as $M$, with size $N\times N$. When there is a connection between two nodes $i$ and $j$, the value ${{e}_{ij}}=1$ is assigned, while ${{e}_{ij}}=0$ otherwise. Note that $M$ is not a symmetric matrix.

In order to optimize the routing of flows, weights ${{w}_{ij}}$ are assigned to individual transmission channels; while ${{w}_{ij}}$ are from the range of natural numbers $(1,v)$ where $v$ can take any value. Weight matrices $W$ have size of $N\times N$.

The weighted adjacency matrix ${M}_{W}$ is determined as the Hadamard product~\cite{bl34} of $M$ and $W$ matrices:

\begin{equation} 
{{M}_{W}}=M\bullet W=\left(\begin{matrix}  {{w}_{11}}\cdot{{e}_{11}}  & \ldots  & {{w}_{1N}}\cdot{{e}_{1N}} \\
   \vdots & \ddots & \vdots \\
   {{w}_{N1}}\cdot{{e}_{N1}} & \cdots & {{w}_{NN}}\cdot{{e}_{NN}} \\
\end{matrix} 
\right) \label{eg_3}
\end{equation}

\begin{figure}[htbp]
\centerline{\includegraphics[scale=1]{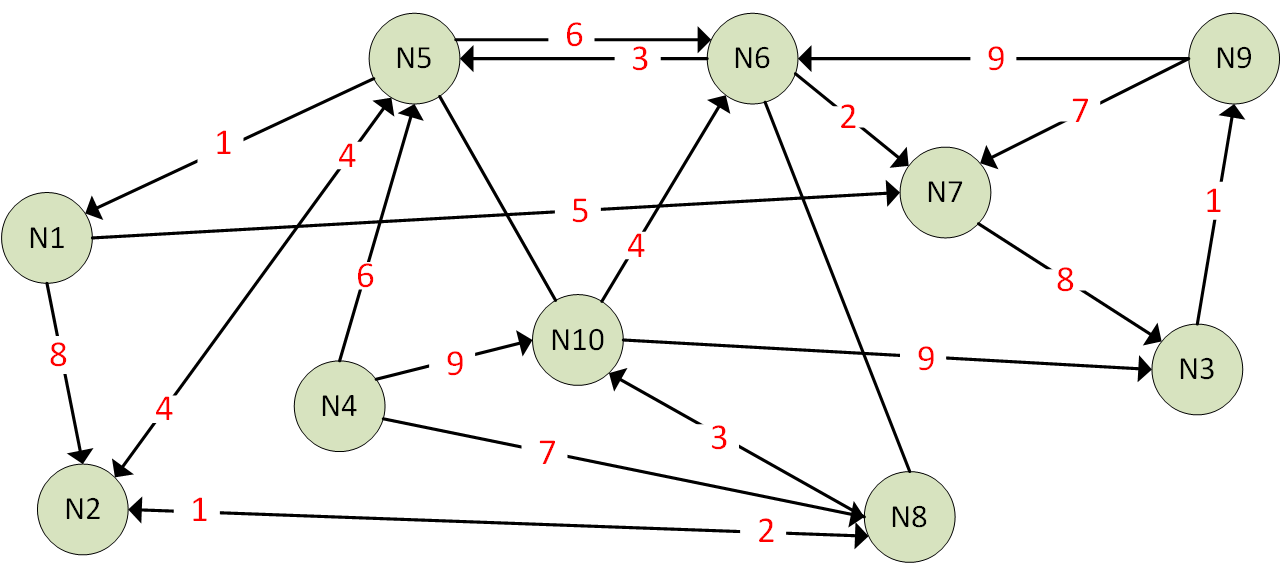}}
\caption{An example graph representation of a weighted, directed interconnection network.}
\label{figure_1}
\end{figure}

Figure~\ref{figure_1} shows a sample graph representation of a network with $N=10$, and $v=9$. For pairs of vertices $(s,d)$, where $s,d\in N$, \textit{homogeneous traffic flow} $f$ specifies requests to transmit information from the source node to the destination node, represented by a flow matrix ${{F}_{sd}}$.
\begin{equation}
    {F_{sd}} = \left( {\begin{array}{*{20}{c}}{{p_{11}}}& \ldots &{{p_{1N}}}\\ \vdots & \ddots & \vdots \\{{p_{N1}}}& \cdots &{{p_{NN}}}\end{array}} \right) \label{eg_4}
\end{equation}
where: $s=d\to {{p}_{sd}}=0$; ${{p}_{sd}}=m \cdot f$, $m \in \mathbb{N}$.
Homogeneous flow $f$ corresponds to the granularity of flows in the network, as is the case with queues, e.g. in the Ethernet network ($f=64\text{kb}$). Total flow ${{p}_{sd}}$, is therefore defined as a multiple of the base flow $f$. When, a unit value $f$ is assumed, then ${{p}_{sd}}=m$. For example, for channels 100 Mb/s and granulation $f=64\text{kb}$, for any edge $\max m=1563$. In what follows, such network will be named a \textit{network with homogeneous flow structure}. The flow matrix can change over the life time of the network. The values of the elements of this matrix can also be predicted in advance~\cite{9163001}.

The matrix $L$ determines the current link load ${{e}_{ij}}$ of the network topology. Here, it is assumed that it will be systematically modified, as a result of the modification of weights in the matrix $W$ and the distribution of flows defined in the matrix ${{F}_{sd}}$.

The path between the vertices $s,d$, for a given flow ${{p}_{sd}}$, will be determined using the Dijkstra algorithm~\cite{bl35}. However, other algorithms can also be used here. Thus, the value for ${{l}_{ij}}$ is determined as the sum of flows ${{p}_{sd}}$ passing through the edge ${{e}_{ij}}$. Taking into account the need of dynamic control of link weights (in the network), the problem of load balancing can be defined, as seeking a set of link weights $W$, for which the maximum number of flows passing through the ``busiest edge'' in the network has been minimized. Therefore, the problem can be formulated as:
\begin{equation}
\min \left( \max \left( {{l}_{ij}} \right) \right) \label{fit_eq}
\end{equation}

Note that, in the algorithm, the load matrix $L$ is represented as a load vector $VL={{[{{l}_{11}},{{l}_{12}},...,{{l}_{1N}},...,{{l}_{N1}},{{l}_{N2}},...,{{l}_{NN}}]}^{T}}$

As mentioned before, the problem of balancing the load of links in the network is NP-hard, i.e. the time to find the optimal solution grows exponentially with the size of the problem. The overriding goal of this contribution is thus to propose a solution characterized by low computational complexity, which will allow actual implementation of the balancing algorithm, and its deployment in a real production system. As a result, the SDN Genetic Algorithm Load Balancer (SDNGALB) is proposed. Let us described it in more details.

\subsection{Operation of the algorithm SDNGALB}
In the first step, the values of the elements ${{w}_{ij}}$ are randomly populated, with natural numbers from the range $(1,v)$. In what follows, $\max v=9$ was assumed. However, for very large networks, with high connectivity, a larger range of weights may be needed. However, this must be determined experimentally, or based on the designers' intuition~\cite{bl33}.

\begin{figure}[htbp]
\centerline{\includegraphics[scale=0.30]{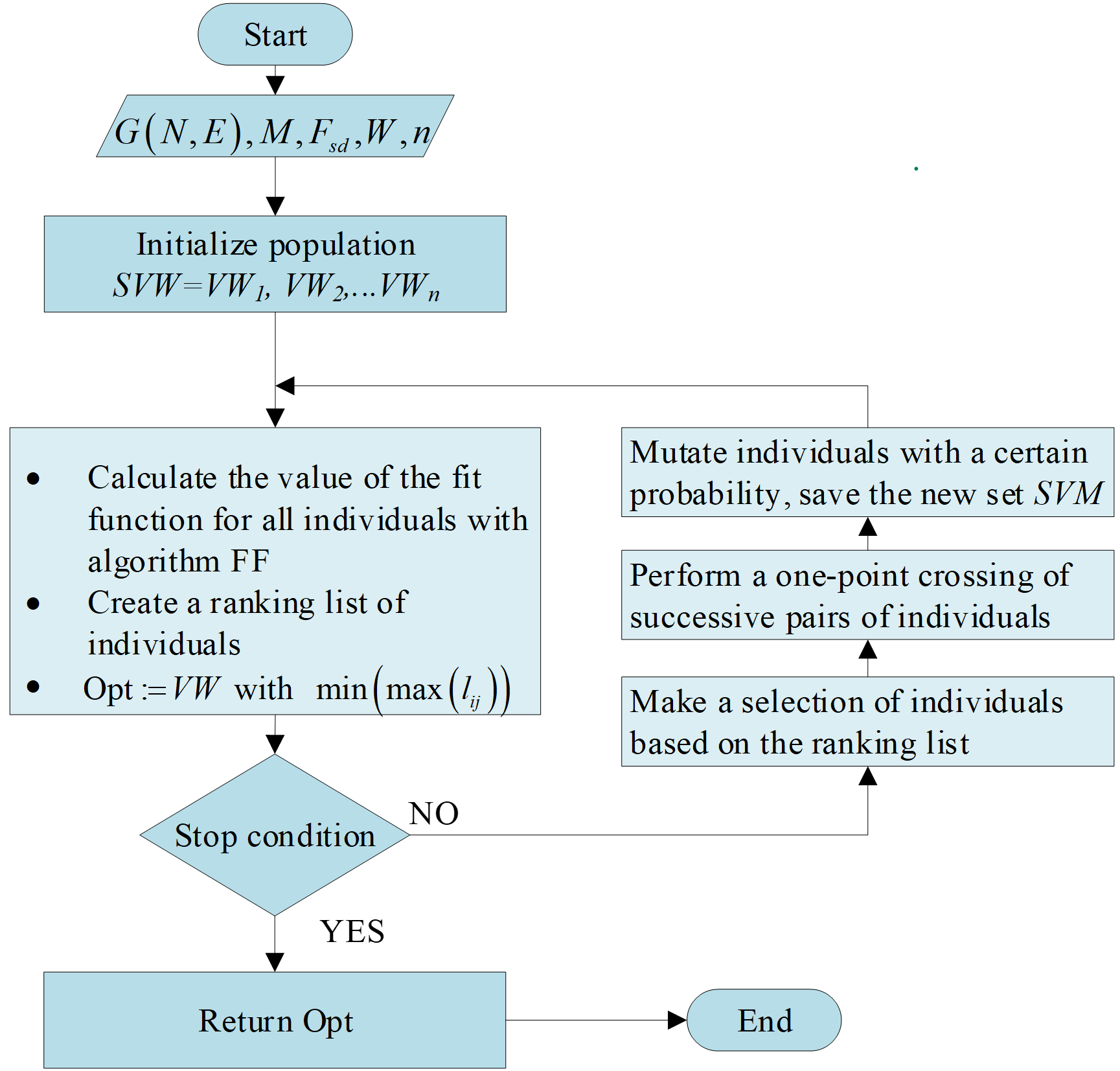}}
\caption{Block diagram of the operation of the SDNGALB algorithm.}
\label{figure_2}
\end{figure}

The genetic algorithm, presented in Figure~\ref{figure_2}, uses standard genetic operators (cross-over and mutation) to process the population of solutions originating from the prior selection. The following decision have been made during the design of SDNGALB (see, also, Figure~\ref{figure_2} for the block diagram of the algorithm).
\begin{enumerate}
    \item The chromosome is the weight list $VW$, of individual network links, created on the basis of the matrix ${{M}_{W}}$, according to the rule: if for any $x,y \in (1,N)$, ${{w}_{xy}}\cdot{{e}_{xy}}=0$ weight should be omitted; if ${{w}_{xy}}\cdot {{e}_{xy}}>0$ the weight should be added to the list (the length of the chromosome is equal to the number of edges in the network graph).
    
    \item The initial chromosomes are randomly generated, from range $(1,v)$. The size of the population is selected experimentally, and is denoted as $n$.
    
    \item The fitness function (FF) is calculated using formula~(\ref{fit_eq}). The FF algorithm is presented, in the form of a pseudocode, as Algorithm~1.

    \item Individuals are ranked on the basis of the value of their fitness function.
    
    \item Pairs of individuals, arranged according to the quality of adaptation, are crossed with each other using the one-point method -- the crossing point is selected randomly, and the parts of the chromosome ``after this point'' are exchanged, creating individuals of a new population.
    
    \item Mutation of individuals occurs with the probability determined by the parameter $mp$, and consists in drawing a new value (from a specified range) of any gene in the chromosome.
    
    \item The optimization process is performed until one of the following stop condition is reached:
    \begin{enumerate}
    
	\item  Stagnation parameter ($sz$) is reached, i.e. number of solutions, during which the obtained optimization result did not improve;
        \item  The maximum number $gs$ of generations is reached.
    \end{enumerate}
\end{enumerate}
The Fitness Function algorithm requires a more detailed explanation. For each of the defined flows ${{F}_{sd}}$, using the Dijkstra algorithm~\cite{bl35}, the shortest paths between the source node and the destination node are calculated (taking into account matrices $M,W,{{M}_{W}}$). Optimization of load distribution in the network is achieved by manipulating the link weights in the matrix $W$, on the basis of which the matrix ${{M}_{W}}$ is built, in a way to eliminate network bottlenecks, by reducing the load on the most frequented edge in the network. For a set of weights $W$, of a given chromosome $VW$, and a set flows ${{F}_{sd}}$, obtained using the Dijkstra algorithm, the routing table $RT$ is determined. It contains a number of rows equal to the number of non-zero matrix elements ${{F}_{sd}}$ and each row contains a list of vertices to be traversed for a given flow ${{p}_{sd}}$, marked as $P({{p}_{sd}})$. For example, a single row in the $RT$ routing table, for a flow between $6$ and $0$ vertices $F_{6,0}$, might look like $P\left( {{F}_{6,0}} \right)\to e_{6,5}, e_{5,3}, e_{3,1}, e_{1,0}$, where $e_{u,d}$ denotes the edge connecting vertices $u$ and $d$. Then, on the basis of $RT$, FF determines the value of the maximum link load in the network $\max \left( {{l}_{ij}} \right)$. 
\par Below is the algorithm for evaluating the value $\max \left( {{l}_{ij}} \right)$ used in the fit function. The source code of the algorithm is available at the website \cite{MB1}\\ \\
\textbf{Algorithm 1} Fit function (FF) \\
\textbf{Data:} $G\left( N,E \right),W,{{F}_{sd}},P\left( {{F}_{sd}} \right)$ \\
\quad 1:\quad	$L:=0,RT:=0$ \\
\quad 2:\quad for each ${{F}_{sd}}$\\
\quad 3:\qquad $P\left( {{F}_{sd}} \right):=Dijkstra\left( G,W,{{F}_{sd}} \right)$ \\
\quad 4:\qquad $RT=RT+P({{F}_{sd}})$ \\
\quad 5:\qquad for each ${{e}_{sd}}\in P\left( {{F}_{sd}} \right)$ \\
\quad 6:\qquad \quad if ${{l}_{sd}}={{e}_{sd}}$ \\
\quad 7:\qquad \quad then ${{l}_{sd}}={{l}_{sd}}+1$ \\
\quad 8:\qquad  end \\
\quad 9:	 end \\
\quad 10:	 return\ $max \left( {{l}_{ij}} \right)$ \\ \\
The results of operation of SDNGALB is the optimal set of link weights $W$, in the given network, on the basis of which a balanced routing table $RT$ is built. This set of weights minimizes the maximum link load in the network. Depending on the specifics of the network, the administrator can establish the time step, at which the algorithm will be used to re-optimize the network. It is also possible to use triggered reconfiguration, e.g. when any of the communication channels reaches a specific throughput limit (e.g. 80\%).

\section{Experimental setup and results \label{section3}}
The code of the algorithm was implemented in Python, during the implementation, apart from the standard Python library modules, also \textit{numpy} and \textit{matplotlib} libraries and the \textit{networkx} package were used. All experiments were carried in a simulation environment with the following parameters: Debian GNU/Linux 10 4.19.0-8-amd64; 8 CPUs Intel® Xeon® CPU E5-2620 v3 @ 2.40GHz; RAM: 39,3 GB. Research, carried in the computer networks laboratory of Rzeszów University of Technology, included implementation of the SDN architecture in the environment of real enterprise class network devices (Extreme and OmniSwitch Alcatel-Lucent), ``bare metal'' switches (Edge Core), and OpenvSwitches. During experiments, a special stand configured for research of phenomena in the ``Internet of Everything'' was used~\cite{stand}. The exact parameters for each experiments will be provided as required.

The first set of tests was focused on the effectiveness of optimization. For flows in the network represented by graph $G\left( N,E \right)$, where $\left| N \right|=10$ and $\left| E \right|=39$, the values of the function (\ref{eg_4}) were compared before and after using the SDNGALB algorithm. Table \ref{tab1} shows the mean value of $\max \left( {{l}_{ij}} \right)$, before and after optimization, calculated as the arithmetic mean of 10 executions of the SDNGALB algorithm for the defined network, for $\left| {{F}_{sd}} \right|=20;30;40;50;100;200$.
\begin{table}[htbp]
\caption{Comparison of the average effectiveness of optimization for a different number of flows in the network.}
\begin{center}
\begin{tabular}{|c|c|c|c|c|c|c|}
\hline
\textbf{Arithmetic mean}&\multicolumn{6}{|c|}{\textbf{$\left| {{F}_{sd}} \right|$}} \\
\cline{2-7} 
\textbf{ of 10 executions} & \textbf{\textit{20}}& \textbf{\textit{30}}& \textbf{\textit{40}} & \textbf{\textit{50}} & \textbf{\textit{100}} & \textbf{\textit{200}} \\
\hline
$\overline{\max \left( {{l}_{ij}} \right)}$ before optimisation &5,6& 8,6 & 8.8 & 11,6 & 21,8 & 41,6\\
\hline
$\overline{\max \left( {{l}_{ij}} \right)}$ after optimisation &2,8& 4 & 4,8 & 5,8& 11,6 & 22,4\\
\hline
\textbf{Effectiveness of optimisation} &50\% & 53\% & 45\% & 50\% & 47\% & 56\%\\
\hline
\end{tabular}
\label{tab1}
\end{center}
\end{table}
As can be seen, application of the SDNGALB algorithm resulted in a noticeable reduction in throughput on the most heavily loaded links in the tested network. Simultaneously, the performance tests showed an increased load on links that were previously not used and those that were only lightly loaded. 

The next set of experiments was performed to compare the time of reaching solution using the genetic algorithm and the exact algorithm searching the entire solution space (i.e. the brute-force search; BF). Obviously, obtained results should be seen only as base-line comparison with the worst case (time-wise), but most efficient (finding best solution) approach. Here, 100 simulations were performed for both algorithms for a network with 4 nodes and 5 edges, and five defined flows. The same optimization result  expressed by the value of the function  (\ref{fit_eq}; the minimum value of the maximum link load in the network) was obtained by both algorithms. For the genetic algorithm the mean time was 0.0118065 seconds, while for the BF algorithm time was 4448.077889 seconds.

In the third set of experiments, the effectiveness of the proposed solution on networks with different topologies was assessed. The speed of obtaining the solution, and the quality of the obtained result, for $\min \left( \max \left( {{l}_{ij}} \right) \right)$ are reported. In the first phase, 1000 simulations were performed, for six different network topologies, in which the initial flows were randomly generated. The characteristics of the networks $Net$, used in the study, is presented in Table \ref{tab2}, where the connectivity parameter $Cn$ should be understood as the percentage ratio of the number of edges in the tested network to the number of edges in the network with all possible connections (represented as a complete graph).
\begin{table}[htbp]
\caption{Parameters of networks used in the research.}
\begin{center}
\begin{tabular}{|c|c|c|c|c|c|c|}
\hline
\textbf{$Net$} & $n4e5$& $n5e11$ & $n6e15$ & $n10e39$ & $n25e219$ & $n50e872$ \\
\hline
 $\left| N \right|$ & 4 & 5 & 6 & 10 & 25 & 50\\
\hline 
 $\left| E \right|$ & 5 & 11 & 15 & 39 & 219 & 872\\
\hline
$Cn$ & 31\% & 44\% & 42\% & 39\% & 35\% & 35\%\\
\hline
  $v$ &1-5& 1-5 & 1-5 & 1-9 & 1-9 & 1-9\\
\hline
$\left| {{F}_{sd}} \right|$ & 5& 10 & 15 & 20 & 45 & 100\\
\hline
\end{tabular}
\label{tab2}
\end{center}
\end{table}

Table~\ref{tab3} presents the average times $Ts$ (measured in seconds) of reaching the solution; for 1000 simulations, performed for each of the networks listed in Table~\ref{tab2}. In this case, the flows were generated randomly for each individual simulation. The algorithm parameters were set as follows: mutation probability: $mp=10\%$; population size: $n=50$; generation size: $gs=500$; stagnation: $sz=100$. Here, as the network size increases, i.e. from 4 nodes and 5 connections to 50 nodes and 872 connections, the solution time remains within an acceptable range. Therefore it can be stipulated that netwrk optimization, based on the proposed SDNGALB algorithm, can find application in real network systems.
\begin{table}[htbp]
\caption{Average result of the load balancing time using the SDNGALB algorithm in seconds.}
\begin{center}
\begin{tabular}{|c|c|c|c|c|c|c|}
\hline
\textbf{$Net$} & $n4e5$& $n5e11$ & $n6e15$ & $n10e39$ & $n25e219$ & $n50e872$ \\
\hline
 $Ts$ & 0.012 & 0.021 & 0.034 & 0.093 & 1.091 & 10.264\\
\hline 
 \end{tabular}
\label{tab3}
\end{center}
\end{table}

\begin{figure}[htbp]
\centerline{\includegraphics[scale=0.4]{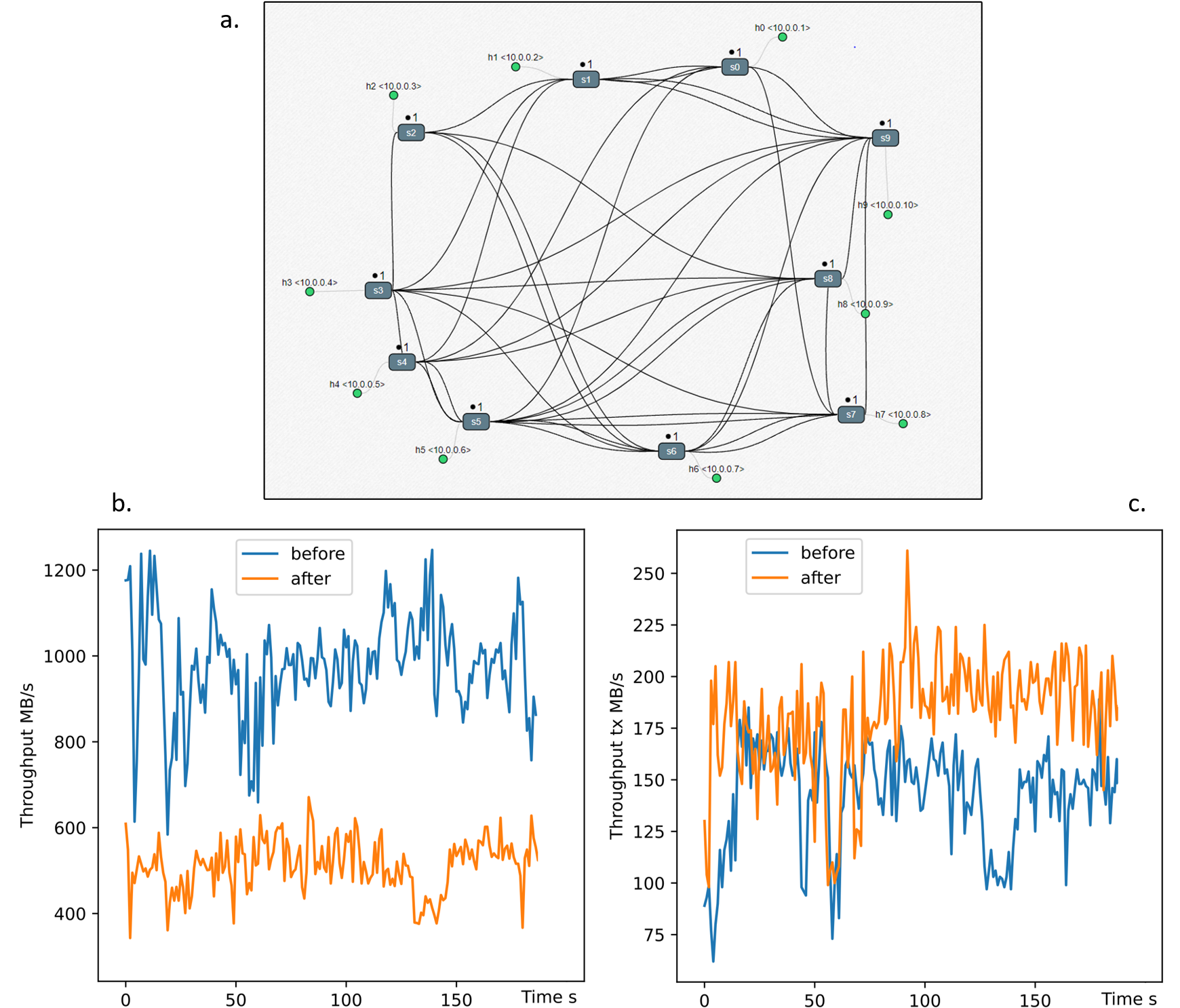}}
\caption{a. The n10e39 network topology used in tests; throughput on links $e_{3,1}$ (b.) and $e_{4,0}$ (c.)  before and after running the SDNGALB}
\label{figure_3}
\end{figure}

The proposed algorithm was implemented in the Mininet environment using OpenvSwitches.  Figure~\ref{figure_3}a shows an example of one the topologies $n10e39$ used during tests. To control the flows, the  scripts written in Python, were used. During the simulation, additional host, named $h0$ to $h9$, was used as a traffic generator.
Application of the SDNGALB algorithm resulted in a noticeable reduction (39\%) in throughput on one of the most heavily loaded links in the network $e_{3,1}$, as shown in Figure~\ref{figure_3}b. Simultaneously, the performance tests showed an increased load on links that were not previously loaded at all and on those that were loaded slightly - see the example load change on link $e_{4,0}$, as shown in the Figure~\ref{figure_3}c.

The final set of experiments compared the performance of the SDNGALB algorithm with two alternative optimizers. The first one -- Ant Colony Optimization Load Balancer (ACOLAB~\cite{bl37}) uses ant colony optimization to determine optimal routes between nodes. The second one -- Dijkstra's Shortest Path Algorithm (DSPA,~\cite{bl38}) is based on the identification of the shortest paths between given nodes, using Djkstra's algorithm, under the assumption that link weights are randomized (using values from a given range) and are not modified later. DSPA did not optimize link weights and served only as a baseline in the context of the execution time.

Experiments were performed on the n10e39 network (with 10 nodes, 39 edges) with the assumption that $\left| {{F}_{sd}} \right|=20$. The performance of the ACOLB algorithm was tested with different parameter values (number of ants: {1, 5, 10, 25, 50, 100, 500}). For each combination of values for ACOLB algorithm, 100 simulations have been run, and the average results are reported. Moreover, simulations were performed with the DSPA. Summary of the averaged results, obtained for all three algorithms, is shown in Table~\ref{tab4}. 
\begin{table}[htbp]
\caption{Summary of averaged results of all tested approaches}
\begin{center}
\begin{tabular}{|c|c|c|c|}
\hline
 & SDNGALB & ACOLB & DSPA \\
\hline
 Execution time in second & 0.09263 & 0,02229 & 0,00546 \\
\hline
 $\overline{\max \left( {{l}_{ij}} \right)}$ & 2,792 & 31,392 & 5,110 \\
\hline 
 \end{tabular}
\label{tab4}
\end{center}
\end{table}

Based on Table \ref{tab4}, it can be concluded that the proposed solution to the problem of optimization of link weights, in the network during routing, makes it possible to achieve nearly twice the lower maximum link load in the network, as compared to the performance of the DSPA algorithm. In contrast, the ACOLB does not deliver satisfactory optimization vis-a-vis the proposed solution.

\section{Concluding remarks \label{section4}}

In this work the need for efficient SDN network flow optimization has been addressed by means of the dedicated genetic algorithm. The overarching goal was to deliver efficient infrastructure for extended cloud infrastructures, where resources, typically realized as services and microservices, can be highly dispersed. The discussed approach is fast, which should allow to quickly modify the routing table, in response to changing traffic patterns. This is particularly important in EC ecosystems, which can be naturally characterized by fast changing communication patterns. Results obtained during tests, completed on actual networks (though in a laboratory) are very encouraging, concerning both the speed and the quality of optimization.

As part of further work, (1) scalability of the proposed algorithm will be tested for large, highly distributed, networks; (2) algorithm will be adapted to modify the physical topology of MESH networks; and, (3) possibility of automatic, adaptive tuning of optimizer parameters, using machine learning techniques will be explored.

\subsubsection{Acknowledgements} Work of Marek Bolanowski and Andrzej Paszkiewicz is financed by the Minister of Education and Science of the Republic of Poland within the “Regional Initiative of Excellence” program for years 2019–2023. Project number 027/RID/2018/19, amount granted 11 999 900 PLN. Work of Maria Ganzha and Marcin Paprzycki was funded in part by the European Commission, under the Horizon Europe project ASSIST-IoT, grant number 957258.

\bibliographystyle{splncs04}
\bibliography{bibliography}

\end{document}